%% file: arxiv.tex
\documentclass[10pt,a4paper]{amsart}
\usepackage[utf8]{inputenc}

\input{settings-arxiv}

\title{Optimizing simultaneous autoscaling for serverless cloud computing}

\author{Harold Ship\textsuperscript{1}}
\author{Evgeny Shindin\textsuperscript{1}}
\author{Chen Wang\textsuperscript{2}}
\author{Diana Arroyo\textsuperscript{3}}
\author{Asser Tantawi\textsuperscript{2}}

\thanks{\textsuperscript{1}IBM Research -- Haifa, Israel}
\thanks{\textsuperscript{2}IBM Research -- Yorktown Heights, NY}
\thanks{\textsuperscript{3}IBM Research -- Almaden, CA}

\date{May 2023}

\begin{document}

\maketitle

\input{src/abstract.tex}

\input{src/introduction.tex}
\input{src/model.tex}
\input{src/experiment.tex}
\input{src/results.tex}
\input{src/relatedwork.tex}
\input{src/conclusion.tex}

\bibliographystyle{IEEEtran}
\bibliography{src/IEEEabrv,src/references.bib}

\end{document}

%% file: settings-arxiv.tex
\usepackage{amsfonts} 
\usepackage{mathtools} 
\usepackage{fixmath} 
\usepackage{relsize} 
\usepackage{bm} 
\usepackage{paralist}

\usepackage{graphicx} 
\graphicspath{{figs/}} 

\usepackage[caption=false,font=footnotesize]{subfig}

\usepackage{url}
\usepackage{hyperref}
\usepackage{tabularx}
\usepackage{etoolbox}
\usepackage{enumitem}
\usepackage{xcolor} 
\usepackage[normalem]{ulem} 

\usepackage{multirow} 
\usepackage{soul} 

\usepackage{array} 
\newcolumntype{H}{>{\setbox0=\hbox\bgroup}c<{\egroup}@{}}

\newtoggle{hidden}

\usepackage{tikz}
\usetikzlibrary{positioning}
\usetikzlibrary{arrows}
\usetikzlibrary{shapes.geometric}
\usetikzlibrary{fit}

\usepackage{etoolbox}

\newcommand{\ignore}[1]{\xspace}

\newbool{IsPrintComment}
\boolfalse{IsPrintComment}

\newcommand{\da}[1]
{
    \ifbool{IsPrintComment}
    {%
        {\bf \color{orange} DA: #1}
    }{}
}

\newcommand{\ES}[1]
{
    \ifbool{IsPrintComment}
    {%
        {\bf \color{blue} ES: #1}
    }{}
}

\newcommand{\hs}[1]
{
    \ifbool{IsPrintComment}
    {%
        {\bf \color{brown} HS: #1}
    }{}
}

\newcommand{\AT}[1]
{
    \ifbool{IsPrintComment}
    {%
        {\bf \color{purple} AT: #1}
    }{}
}

%% file: src/abstract.tex
\begin{abstract}
This paper explores resource allocation in serverless cloud computing platforms and proposes an optimization approach for autoscaling systems. Serverless computing relieves users from resource management tasks, enabling focus on application functions. However, dynamic resource allocation and function replication based on changing loads remain crucial. Typically, autoscalers in these platforms utilize threshold-based mechanisms to adjust function replicas independently.

We model applications as interconnected graphs of functions, where requests probabilistically traverse the graph, triggering associated function execution. Our objective is to develop a control policy that optimally allocates resources on servers, minimizing failed requests and response time in reaction to load changes. Using a fluid approximation model and Separated Continuous Linear Programming (SCLP), we derive an optimal control policy that determines the number of resources per replica and the required number of replicas over time.

We evaluate our approach using a simulation framework built with Python and simpy. Comparing against threshold-based autoscaling, our approach demonstrates significant improvements in average response times and failed requests, ranging from 15\% to over 300\% in most cases. We also explore the impact of system and workload parameters on performance, providing insights into the behavior of our optimization approach under different conditions. Overall, our study contributes to advancing resource allocation strategies, enhancing efficiency and reliability in serverless cloud computing platforms.

\end{abstract}

%% file: src/introduction.tex
\section{Introduction}\label{sec:introduction}
The serverless paradigm has gained popularity due to its benefits such as  elimination of infrastructure management, cost efficiency, automatic scaling, faster time to market, compatibility with micro services architecture, and suitability for event-driven and real-time processing. Serverless computing allows users to develop and deploy code without the need to manage the underlying infrastructure, and it employs a pay-as-you-go model where users are charged based on actual resource consumption. A key feature of this computing model is autoscaling, a dynamic process that adjusts the allocation of computational resources in response to the demand on a function. This autoscaling mechanism monitors function invocations, automatically increasing function instances to meet high demand and decreasing them when demand subsides, optimizing resource use and cost.

Various approaches have recently been proposed to address some of the existing challenges in serveless platforms, such as function request scheduling and resource allocation, using carefully designed heuristics \cite{Kaffes2021, Suresh2020, Tariq2020, Wen2022, Yang2022}. Some approaches (\cite{Tariq2020, Wen2022}) allow the definition of function chains while others (\cite{Kaffes2021, Suresh2020}) use queueing theory to define optimal regimes for single functions. Unfortunately, none of the aforementioned solutions consider the application of queueing theory to the function chains. 



In this paper, we propose a novel proactive scheduling policy based on the Multi-class Queueing Network model (MCQN). The MCQN model offers several important benefits:

\begin{compactitem}
\item It enables the definition of different types of resources (e.g., CPU, GPU, RAM) with finite resource capacities.
\item It allows for the definition of various types of functions based on load patterns, service durations, resource demands, and other relevant factors.
\item It supports the specification of function chains, wherein each request, upon being serviced, can spawn additional requests that need to be served by other serverless functions.
\item It takes into account the actual state of the network, without assuming that the network state is steady.
\item Existing algorithms allow for the determination of asymptotically optimal control policies for very large networks within a reasonable time frame.
\end{compactitem}

To find the optimal control policy for the MCQN model, we employ a fluid approximation approach that enables the determination of an asymptotically optimal control policy by solving a mathematical optimization problem known as the Separated Continuous Linear Programming problem (SCLP). One key feature of the SCLP is the existence of the Revised SCLP-Simplex algorithm, which allows for the solution of very large problems \cite{Shindin2021}. This capability enables the recomputation of the optimal policy at a desired frequency, thus adapting the policy to changes in function demand. Additionally, we have devised a method to translate the SCLP solution into an optimal policy that dynamically adjusts the number of resources per replica and the number of replicas over time, which we refer to as the fluid policy.

To evaluate the effectiveness of our proposed approach, we have developed a simulation model and conducted a numerical study comparing it against a threshold-based autoscaler. In this study, we analyze the impact of various network parameters, such as network size, desired request timeout, and function heterogeneity, on performance measures including average response times and the number of failed requests under both autoscaling and fluid policies.


%

%% file: src/model.tex
\section{The Model}\label{sec:model}
In this section, we introduce a comprehensive modeling framework for dynamic resource allocation in the serverless ecosystem. We first describe the
model for a simple network that adopts paradigm considered by Harrison and
Wein \cite{Harrison1989} and then describe the general problem formulation.
\setlength\abovedisplayskip{4pt}
\setlength\belowdisplayskip{4pt}
\subsection{Simple network}
\label{sec:simple_model}
\begin{figure}[htbp]
    \centering
    \resizebox{\linewidth}{!}{
    \begin{tikzpicture}
    
        \node[rectangle, draw=blue, minimum height=2cm, minimum width=1cm] (B1) at (0, 0) {$LB_1$};
        \node[rectangle, draw=blue, below=4cm of B1, minimum height=2cm, minimum width=1cm] (B2) {$LB_2$};
        \node[rectangle, draw=blue, right=4.5cm of B2, minimum height=2cm, minimum width=1cm] (B3) {$LB_3$};
        
        \path[below left,shorten >=1pt,draw] (-3, 0) edge[->] node {$\mathcal{A}_1(t)$} (B1);
        \path[below left,shorten >=1pt,draw] (-3,-6) edge[->] node {$\mathcal{A}_2(t)$} (B2);
        
        \node[ellipse, draw=brown, label=below:$\vdots$, minimum size=1cm] (R_1_1) at (1.5, 1) {$R_{1}^{1}$};
        \node[ellipse, draw=brown, minimum size=1cm, below=of R_1_1] (R_1_R) {$R_{1}^{r_1}$};

        \node[ellipse, draw=brown, label=below:$\vdots$, minimum size=1cm] (R_2_1) at (1.5, -5) {$R_{2}^{1}$};
        \node[ellipse, draw=brown, minimum size=1cm, below=of R_2_1] (R_2_R) {$R_{2}^{r_2}$};

        \node[ellipse, draw=brown, label=below:$\vdots$, minimum size=1cm] (R_3_1) at (7, -5) {$R_{3}^{1}$};
        \node[ellipse, draw=brown, minimum size=1cm, below=of R_3_1] (R_3_R) {$R_{3}^{r_3}$};
        
        \node[rectangle, draw=red, inner sep=0.5cm, text height=3cm, anchor=north, label={[below]above:Function 1}, fit=(B1)(R_1_1)(R_1_R)] (S1) {};
        \node[rectangle, draw=red, inner sep=0.5cm, text height=3cm, anchor=north, label={[below]above:Function 2}, fit=(B2)(R_2_1)(R_2_R)] (S2) {};
        \node[rectangle, draw=red, inner sep=0.5cm, text height=3cm, anchor=north, label={[below]above:Function 3}, fit=(B3)(R_3_1)(R_3_R)] (S3) {};
        
        \node[rectangle, draw=black, label={below:Server 1}, fit=(S1)(S2), inner sep=0.5cm] at (1, -3) (Server_1) {};
        \node[rectangle, draw=black, label={below:Server 2}, fit=(S3), inner sep=0.5cm] at (6.5, -6) (Server_2) {};
        
        \path[below,shorten >=1pt,draw] (R_1_1) -- (3,1) -- (3,0) edge[->] node {$\mathcal{D}_1(t)$} (10,0);
        \path[below,shorten >=1pt,draw] (R_1_R) -- (3,-1) -- (3,0);

        \path[below,shorten >=1pt,draw,text width=1.3cm] (R_2_1) -- (3,-5) -- (3,-6) edge[->] node {$\mathcal{D}_2(t)$} (B3);
        \path[below,shorten >=1pt,draw] (R_2_R) -- (3,-7) -- (3,-6);

        \path[below right,shorten >=1pt,draw] (R_3_1) -- (8.25,-5) -- (8.25,-6) edge[->] node {$\mathcal{D}_3(t)$} (10,-6);
        \path[below,shorten >=1pt,draw] (R_3_R) -- (8.25,-7) -- (8.25,-6);

    \end{tikzpicture}
    }
    \caption[Simple network example]{Simple network example.}
    \label{fig:simulator-architecture}
\end{figure}
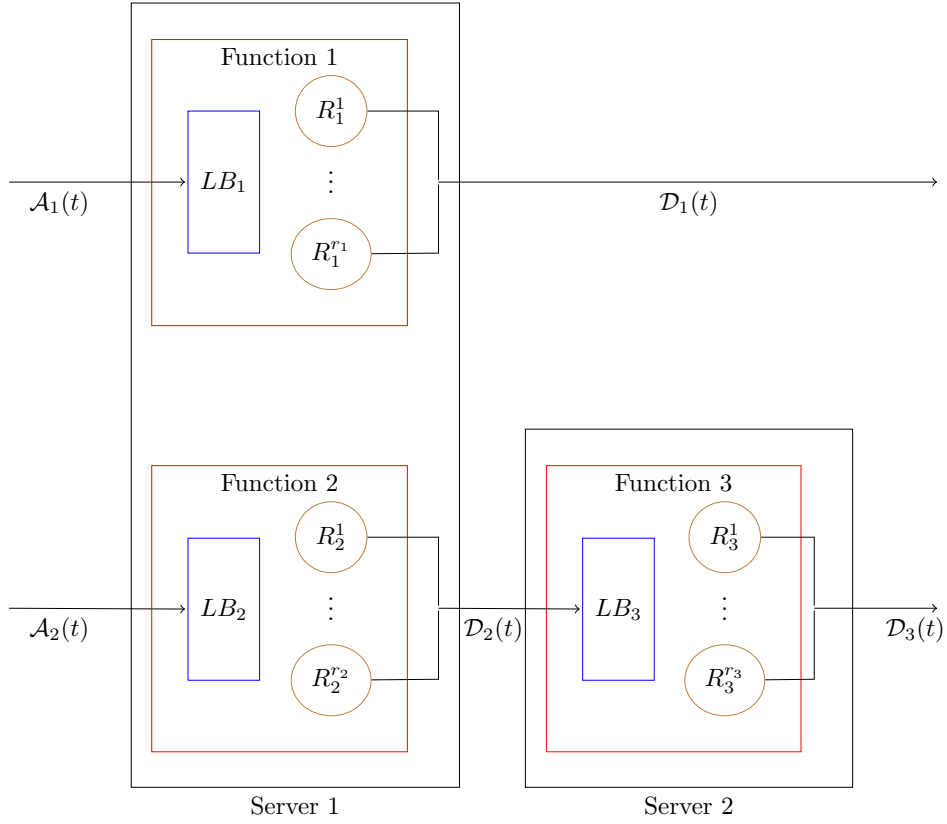
The diagram shown in Figure \ref{fig:simulator-architecture} illustrates a simplified model addressing the serverless resource allocation problem. 
This model involves a network consisting of two servers, labeled as $i=1,2$, and three services or functions denoted as $k=1,2,3$. The model considers distinct allocations of functions to the servers. Notably, functions 1 and 2 can exclusively be allocated to server 1, while function 3 can solely be assigned to server 2. This model is commonly referred to as a criss-cross network.

In order to apply this model to the serverless ecosystem, we consider that each function is associated with multiple replicas, denoted as $R_k^{l}$, where $l$ varies from 0 to $r_k$. The value of $r_k$ may change dynamically over time. It is assumed that each replica requires a specific quantity of CPU resources denoted as $d_k$, while each server possesses a CPU capacity of $b_i$ units. Requests for functions 1 and 2 originate from external sources, while requests for function 3 are generated by function 2.

The arrival of requests adheres to a homogeneous Poisson process, denoted as $\mathcal{A}_k(\lambda_k)$, with the exception of function 3 where $\lambda_3 = 0$, signifying the absence of exogenous requests for this function. The load balancer is responsible for uniformly distributing incoming requests among the allocated replicas. Each replica is capable of accommodating a maximum of $y_k$ concurrent requests. When the load balancer assigns a request to a replica that has reached its maximum capacity, the request fails to be processed.

Service times are exponentially distributed, with a rate denoted as $u_j(t)$, which is proportional to the number of allocated CPU units, $n_j$. The objective of this model is to identify a policy that determines the optimal allocation of CPU resources per replica and the optimal number of replicas over a predetermined time horizon $T$. This policy aims to minimize the number of failed requests and waiting times within the system.

The research presented in this paper adopts a fluid approximation model, which offers a distinct perspective by representing the system as a fluid rather than focusing on individual requests. This fluid model considers the aggregation of CPUs assigned to specific types of replicas, with $\eta_k(t)$ denoting the total number of CPUs allocated to function $k$.

To estimate the total number of concurrent requests for function $k$ at a given time $t$, the fluid approximation model introduces the concept of fluid in the buffer, represented as $x_k(t)$. The dynamics of the system can be described by the following equations:
\begin{equation}
\label{eqn:dynamic}
\begin{aligned}
x_1(t) &= \alpha_1 + \lambda_1 t - \int_0^t \mu_1(s) \eta_1(s) ds \\
x_2(t) &= \alpha_2 + \lambda_2 t - \int_0^t \mu_2(s) \eta_2(s) ds \\
x_3(t) &= \int_0^t \mu_2(s) \eta_2(s) - \mu_3(s) \eta_3(s) ds
\end{aligned}
\end{equation}

Here, $\alpha_1$ and $\alpha_2$ represent the initial amount of fluid in the corresponding buffers, while $\mu_j$ denotes the processing rates per unit of fluid per single CPU. The model also imposes constraints on the number of allocated CPUs:
\begin{equation}
\label{eqn:resource_cap}
\eta_1(t) + \eta_2(t) \leq b_1 \quad \text{and} \quad \eta_3(t) \leq b_2
\end{equation}
These constraints ensure that the number of allocated CPUs does not exceed the capacity of the respective servers, which are denoted as $b_1$ and $b_2$, respectively.

It is worth noting that the fluid approximation model does not inherently account for failed requests, which means that it may assign zero CPUs to certain functions during specific time intervals, even if the corresponding buffer is not empty. To address this issue and prevent such behavior, we introduce lower bounds on the number of allocated CPUs by enforcing $\eta_k(t) \geq 1$. By doing so, we ensure that for each function is allocated at least one CPU, thereby mitigating the possibility of completely disregarding a function's resource requirements during certain time intervals.

The primary objective of the fluid approximation model is to minimize the overall amount of fluid present within the system over a planned time horizon. This objective leads to the formulation of an optimization problem:
\begin{equation}
\begin{aligned}
\min & \int_0^T \sum_{k=1}^K x_k(t) dt \\
\text{s.t.} & \text{Constraints } (\ref{eqn:dynamic}, \ref{eqn:resource_cap}) \\
& x(t) \geq 0, \eta(t) \geq 1
\end{aligned}
\end{equation}

The solution to this problem involves a piecewise constant control $\eta(t)$, allowing the allocation of CPUs obtained as part of the solution to be divided into the amount of CPU per replica and the number of replicas per time period. 

\subsection{General model}
\label{sec:gen_model}
The general model enables the allocation of $K$ different functions to $I$ servers, allowing for each function to be assigned to one or more servers. The allocations are defined by $j=(k, i)$, representing the possible combinations, resulting in a total of $J$ allocations. A unique allocation is achieved when $j=k$. There are $M$ distinct resources, such as CPU, RAM, GPU, and others with each server having a capacity of $b_i^m$ resources. Requests for function $k$ can originate from external sources or other functions.

Exogenous requests for service $k$ follow a Poisson process $\mathcal{A}_k(\lambda_k)$ with an average arrival rate of $\lambda_k$ requests per unit time. Requests of type $k'$ processed by server $i'$ either generate requests of another type $k$ with probability $p_{j=(k',i'),k}$ or exit the system with a probability of $1-\sum_k p_{j,k}$.

Each allocated function possesses a number of replicas denoted as $R_j^{l}$, where $l$ ranges from 0 to $r_j(t)$. The value of $r_j(t)$ may change dynamically over time. It is assumed that each replica requires a specific amount of resources of type $m$, denoted as $d_j^m$. The load balancer ensures that incoming requests are uniformly distributed among the allocated replicas, even if they are on different servers.

Service times are exponentially distributed with a rate denoted as $u_j(t)$. This rate depends on the allocated resources, such that $u_j(t)=\min_m{g_j^m(r_j(t) n^m_j)}$, where $g_j^m(\cdot)$ represents the dependency of the processing rate on the allocated resources of type $m$. It is assumed that $g_j^m(\cdot)$ are concave functions.

The objective of this model is to identify a policy that optimally allocates resources per replica and determines the optimal number of replicas within a fixed time interval $[0,T]$.

Similar to the example described in Section \ref{sec:simple_model}, we adopt a fluid approximation model where $K$ types of fluids are served by $I$ servers. There are $J$ flows, where $f(j) = k$ if flow $j$ empties buffer $k$, and $s(j) = i$ if flow $j$ is served by server $i$. Flows that empty the same buffer can only be served by different servers, so if $f(j') = f(j'')$, then $s(j') \ne s(j'')$. Flow $j'$ transfers processed fluid from buffer $f(j')$ to other buffers according to proportions $p_{j',k}$ or leaves the system in proportion $1-\sum_k p_{j',k}$. The service rate of flow $j'$ depends on the number of allocated resources, such that $u_{j'}(t)=\min_m{g_{j'}^m(\eta_{j'}^m(t)}$, where $\eta_{j'}^m(t)$ represents the amount of resource $m$ assigned by server $s(j')$ to flow $j'$. The concave functions $g_j^m(\cdot)$ can be approximated by piecewise linear functions $g_j^m(\eta_j^m) = \sum_{\ell=1}^{L^m_j} \mu_{j,\ell}^m \eta_{j,\ell}^m$.

The dynamics of the system can be described by the following equations:
\begin{equation}
\label{eqn:gen_dynamic}
\begin{array}{l}
\displaystyle x_k(t) = \alpha_k + \lambda_k t - \sum_{j: f(j)=k} \int_0^t u_j(s) ds \\
\displaystyle + \sum_{j: f(j)\ne k} \int_0^t p_{j,k} u_j(s) ds \quad k=1,\dots,K, t \in [0,T]
\end{array}
\end{equation}
The service rates are subject to the following constraints:
\begin{equation}
\label{eqn:gen_proc_limit}
u_j(t) \le \sum_{\ell+1}^{L^m_j} \mu^m_{j,\ell} \eta^m_{j,\ell}(t), \quad \forall j,m, t \in [0,T]
\end{equation}
where the number of resources of type $m$ assigned to serve flow $j$ at time $t$ is $\sum_{\ell+1}^{L^m_j} \eta^m_{j,\ell}(t)$.

The number of resources is limited by the total amount of resources available at each server, resulting in resource utilization constraints:
\begin{equation}
\label{eqn:gen_resource_cap}
\sum_{j:s(j)=i}\sum_{\ell+1}^{L^m_j} \eta^m_{j,\ell}(t) \le b^m_i, \quad \forall i,m, t \in [0,T]
\end{equation}

Within the fluid approach, it is also possible to model the quality of service by ensuring that all fluid arriving in buffer $k$ is processed within a specified time $\tau_k$. This leads to the constraint $x_k(t) \le \sum_{j: f(j)=k} \int_0^t u_j(s) ds$. When there are no endogenous inflows to buffer $k$, this constraint can be reformulated as:
\begin{equation}
\label{eqn:gen_timeout}
x_k(t) \le \lambda_k \tau_k \quad \forall t \in [0,T]
\end{equation}
which is equivalent to an upper bound on the buffer size. If there are endogenous inflows to buffer $k$, the system can be remodeled by considering a separate buffer for each inflow.

Additionally, different priorities or weights for different serverless functions can be modeled by introducing a cost $c_k$ for each function. The primary objective of the fluid approximation model is to minimize the overall amount of fluid present within the system over a planned time horizon. This objective leads to the formulation of an optimization problem:
\begin{equation}
\label{eqn:gen_model}
\begin{aligned}
\min & \int_0^T \sum_{k=1}^K c_k x_k(t) dt \\
\text{s.t.} & \text{Constraints } (\ref{eqn:gen_dynamic}, \ref{eqn:gen_proc_limit}, \ref{eqn:gen_resource_cap}, \ref{eqn:gen_timeout}) \\
& x(t), \eta(t) \geq 0
\end{aligned}
\end{equation}

Problem (\ref{eqn:gen_model}) can be classified as a special case of the SCLP problem. This problem exhibits optimal controls $\eta(t)$, that are piecewise constant, with a bounded number of intervals ranging from $n=1$ to $N$ (see \cite{Weiss2008}). To determine the optimal control in terms of the number of replicas and the resources assigned per replica, the following optimization problem can be considered:
\begin{equation}
\label{eqn:gen_discr_model}
\begin{aligned}
\min_{d_j^m, r_{j,n}} & \sum_{n=1}^N \sum_{m=1}^M \sum_{j=1}^J \tau_n w_m d_j^m r_{j,n} \\
\text{s.t.} & d_j^m r_{j,n} \ge \eta^m_{j,n} \\
& \sum_{s(j)=i} d_j^m r_{j,n} \le b^m_{i} \\
& d_j^m \geq \underline{d}_j^m, r_{j,n} \in \mathbb{N}
\end{aligned}
\end{equation}

In this optimization problem, $N$ represents the number of intervals in the SCLP solution, $\tau_n$ denotes the lengths of interval $n$, $\eta^m_{j,n} = \eta^m_j(t)$ for $t \in [t_{n-1}, t_n)$ represents the amount of resource $m$ assigned to flow $j$ according to the optimal solution of SCLP. The parameter $w_m$ indicates the importance of resource $m$, while $b^m_{i}$ represents the amount of resources of type $m$ available on server $i$. The value $\underline{d}_j^m$ denotes the minimum reasonable amount of resources of type $m$ per replica of type $j$. The decision variables in this problem are $d_j^m$, which represents the amount of resources of type $m$ to be assigned to a replica of type $j$, and $r_{j,n}$, which denotes the number of replicas of type $j$ during the time interval $[t_{n-1}, t_n)$.

It is important to note that problem (\ref{eqn:gen_discr_model}) can be viewed as a constraint satisfaction problem. As a result, there are several potential approaches that can be utilized to discover a feasible solution for this problem. For example, one can find optimal solution for the longest time interval and then use obtained values of resources allocated per replica to determine optimal number of replicas for all intervals. 

%% file: src/experiment.tex
\section{Simulations and Performance measures}\label{sec:experiment}
This section presents the simulation model, and defines various performance measures, that will be used to compare our approach with the standard autoscaler across different network configurations.

\subsection{Simulation Model}\label{sec:experiment:simulations}

The simulator is written in Python using the simpy package.

\begin{enumerate}


\item{Simulating arrivals}

As discussed in section \ref{sec:model}, function call requests arrive according to a Poisson process. To simulate these arrivals, we utilize a single Poisson process denoted as $\mathcal{A}(\sum_{k=1}^K \lambda_k)$. In this simulation, the arrival time for the next request is randomly generated from an exponential distribution with a parameter of $\sum_{k=1}^K \lambda_k$. Once the arrival time is determined, the type of request is selected by drawing from a multinomial distribution with parameters $\frac{\lambda_k}{\sum_{k=1}^K \lambda_k}$ for $k=1,\dots,K$. This simulation approach is equivalent to simulating $K$ independent Poisson processes denoted as $\mathcal{A}(\lambda_k)$.
The arrivals are sent to the load balancer for the type.

\item{Resources}

for simulation purpose we consider only CPU utilization. Other resources discussed in section \ref{sec:gen_model} are not considered. We assume that each replica requires exactly $1$ CPU. In cases where the control policy assigns a different number of CPUs, the number is rounded up to the nearest integer, and the corresponding integer value is used as the number of replicas. 

\item{Load balancing}

The load balancer assigns incoming requests to replicas using a round-robin policy.

\item{Concurrency}

Concurrency is implemented by fixed-size queue, where the size is equal to the maximum number of concurrent requests per replica.
This means that any requests that arrive when CPU is busy are added to the queue, as long as the queue is not at maximum capacity.
This has the effect of ``slowing down'' the system as the number of concurrent requests increases.

\item{Processing}

Requests are served based on the First-Come-First-Served (FCFS) policy. Service times are drawn from the exponential distribution with rate $\mu_j$. 

\item{Control policies}

We consider two types of control policies. The first type is based on an auto-scaling approach\AT{Do we need a reference here? \cite{}}that enables scaling the number of replicas of a service, both up and down, based on the failure of the load balancer to find a free replica or the detection of a replica with no requests, respectively. This policy requires specifying an initial, minimum, and maximum number of replicas.

The second type of policy is based on the fluid approximation model discussed in section \ref{sec:model}. For this policy, we have a two-dimensional matrix representing the number of replicas for each function at different time intervals, along with a vector specifying the lengths of the intervals.

\end{enumerate}

\subsection{Performance measures}
\label{sec:experiments:kpis}

\paragraph{Holding cost}

The holding cost is the cost of items waiting to be serviced. We compute it by multiplying the unit service cost by the sojourn time of all requests that arrive at a buffer.
The sojourn time of a request is computed as the arrival time of the request subtracted from one of the following: i) the completion time, for requests that have completed successfully, ii) the time the request is removed from the queue for requests that time out before starting, iii) the end-of-interval time for requests that are still in the queue when the simulation interval ends.
Requests that fail to find a replica upon arrival do not count towards the holding cost.

\paragraph{Average response time}

The average response time is the average time that a request is in the system.
It is computed as the average of the difference between the completion time and the arrival time for all requests that successfully reach completion.

\paragraph{Failures and timeouts}

We count the failures and timeouts of the simulation.
Failure occurs when a request cannot find an free replica when it arrives.
Timeout occurs when a request has spent more than the pre-defined timeout period in the queue.

%% file: src/results.tex
\section{Experiments and Results}\label{sec:results}
To assess the effectiveness of our approach, we conduct a series of experiments that compare it to the standard autoscaler in terms of key performance indicators (KPIs) discussed in Section \ref{sec:experiments:kpis}, across various network configurations.

\subsection{Experimental setup}\label{sec:setup}
In the following experiments the solutions of SCLP for the fluid model were obtained using the Revised SCLP-simplex algorithm described in \cite{Shindin2021} and implemented in Python.\AT{Do we need a reference here? \cite{}}The computation time to obtain the optimal solution on a laptop computer ranged from less than 1 second to 25 seconds, depending on the problem size. Additional results on computational time for similar problems can be found in \cite{Shindin2021}. Instead of solving problem \ref{eqn:gen_discr_model}, we assumed that each replica utilizes exactly $1$ CPU, and other resources were not considered. Therefore, the number of replicas in the fluid model was considered to be equal to $\lceil \eta(t) \rceil$.

In Section \ref{sec:comparison_small}, we present an initial comparison between the fluid policy and the autoscaler, focusing on the arrival and departure processes of individual functions within the criss-cross network discussed in Section \ref{sec:simple_model}. The purpose of this experiment is to examine the performance of these policies at a granular level.

In Sections \ref{sec:comparison_size}-\ref{sec:comparison_hetero}, we analyze the effect of a single parameter on the performance of a queuing network. For each experiment, we utilized the unique allocation model with a base example of a network consisting of 10 servers, each handling 5 function types. Unless otherwise stated, the requests arrived according to a Poisson process with a mean of 100, and the service times followed an exponential distribution with a mean of $1/2.1$ time units.

Each server had a maximum capacity of $250$ replicas, distributed among the 5 function types. At the initial time ($t=0$), the system was initialized with $100$ calls of each function type. By default, each replica of the service could handle up to 100 concurrent requests. For autoscaling, the maximum number of replicas for each function type was set as $250$ divided by the number of function types, resulting in $50$ replicas for each function type. The initial number of replicas was set to $10\%$ of the maximum number.

The reported results represent the average values obtained from 100 simulations of the queue for each configuration.



\subsection{Comparison on a small network}
\label{sec:comparison_small}
\begin{figure*}[!t]
\centering
    \subfloat[Service 1, autoscaling]{\includegraphics[width=0.32\textwidth]{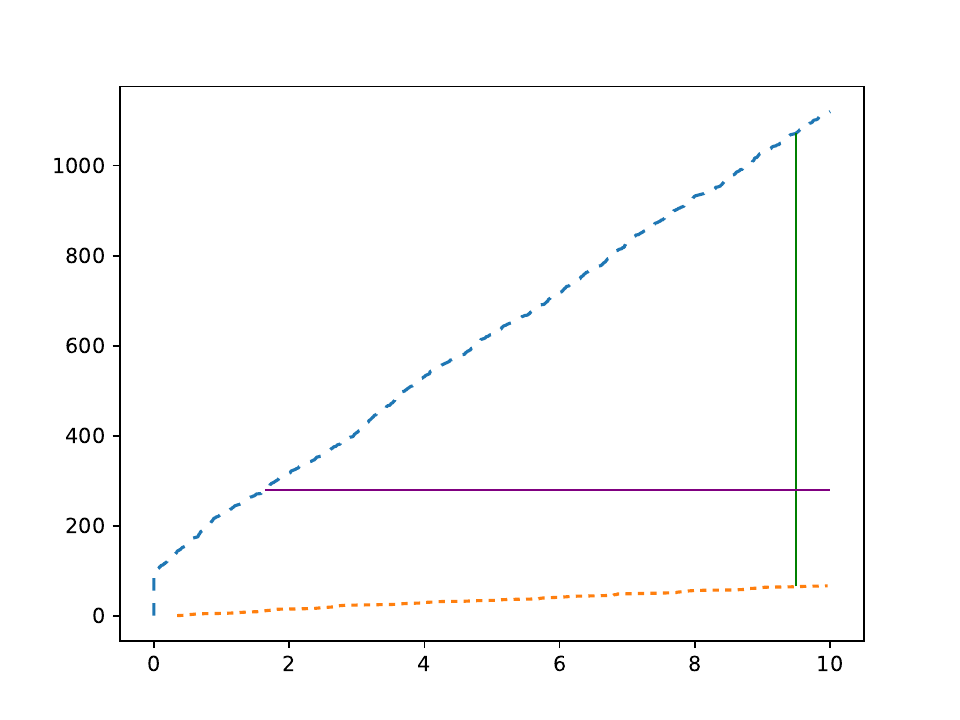}
    \label{fig:criss_cross_cumulative:s1:auto}}
    \subfloat[Service 2, autoscaling]{\includegraphics[width=0.32\textwidth]{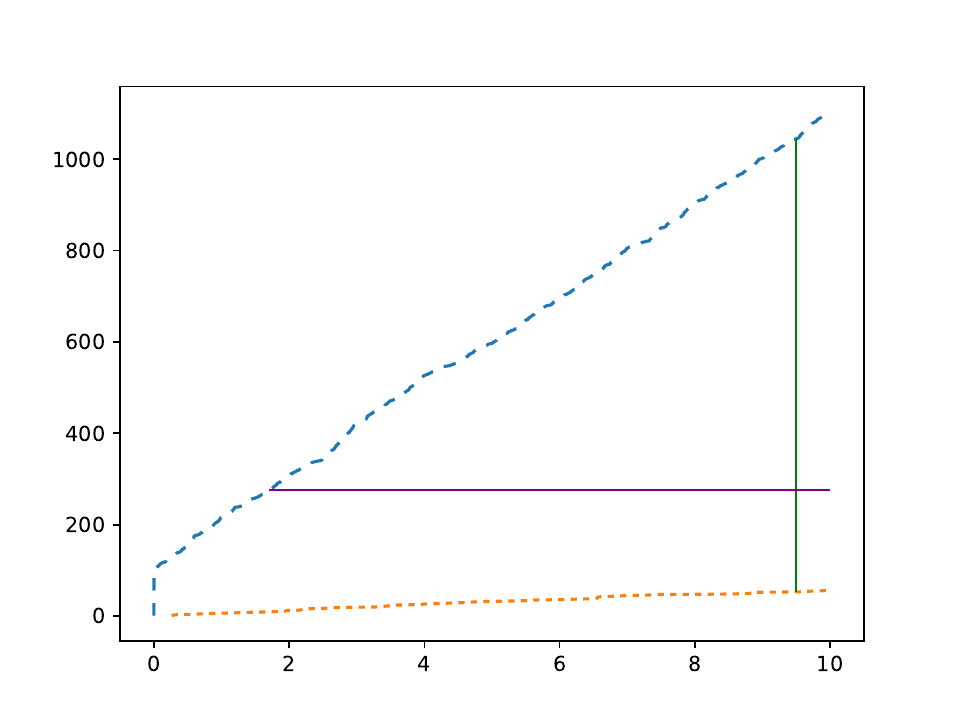}
    \label{fig:criss_cross_cumulative:s2:auto}}
    \subfloat[Service 3, autoscaling]{\includegraphics[width=0.32\textwidth]{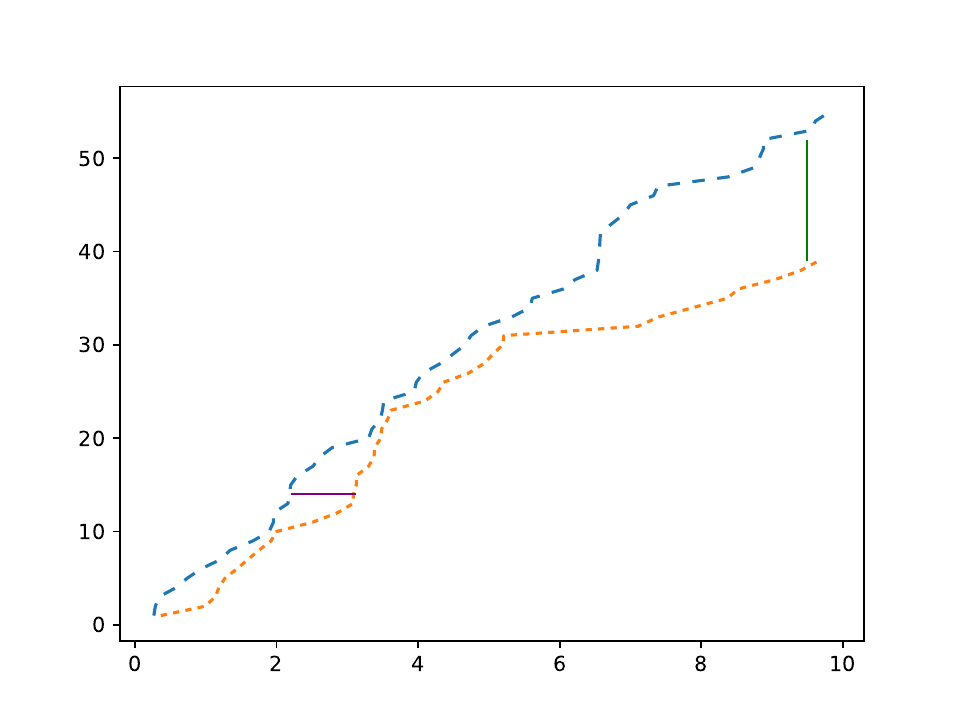}
    \label{fig:criss_cross_cumulative:s3:auto}}
    \hfil
    
    \subfloat[Service 1, fluid]{\includegraphics[width=0.32\textwidth]{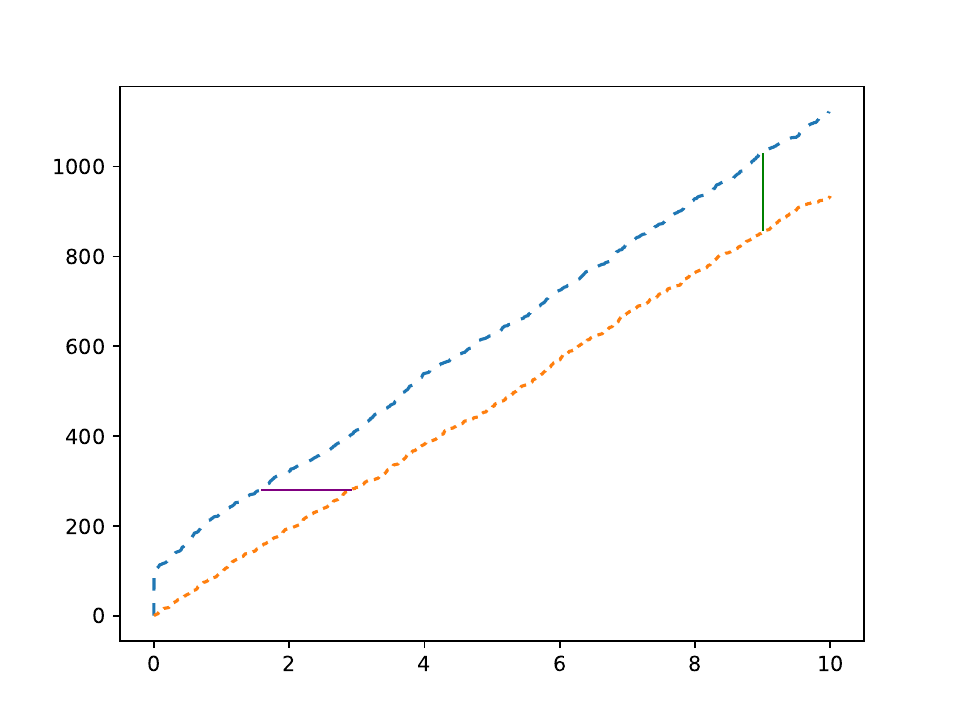}
    \label{fig:criss_cross_cumulative:s1:sclp}}
    \subfloat[Service 2, fluid]{\includegraphics[width=0.32\textwidth]{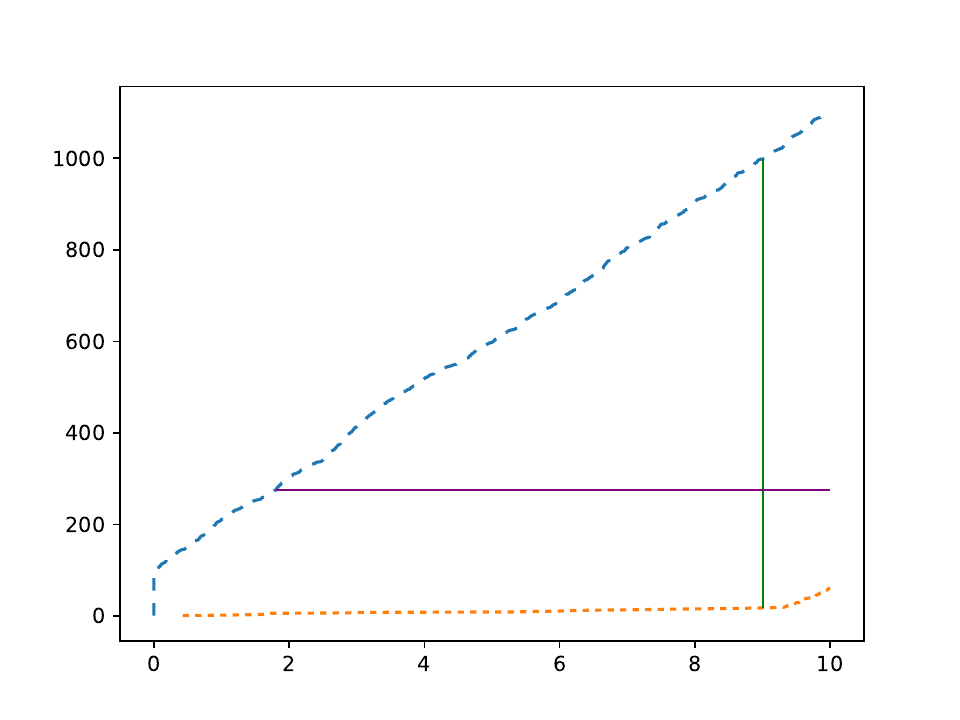}
    \label{fig:criss_cross_cumulative:s2:sclp}}
    \subfloat[Service 3, fluid]{\includegraphics[width=0.32\textwidth]{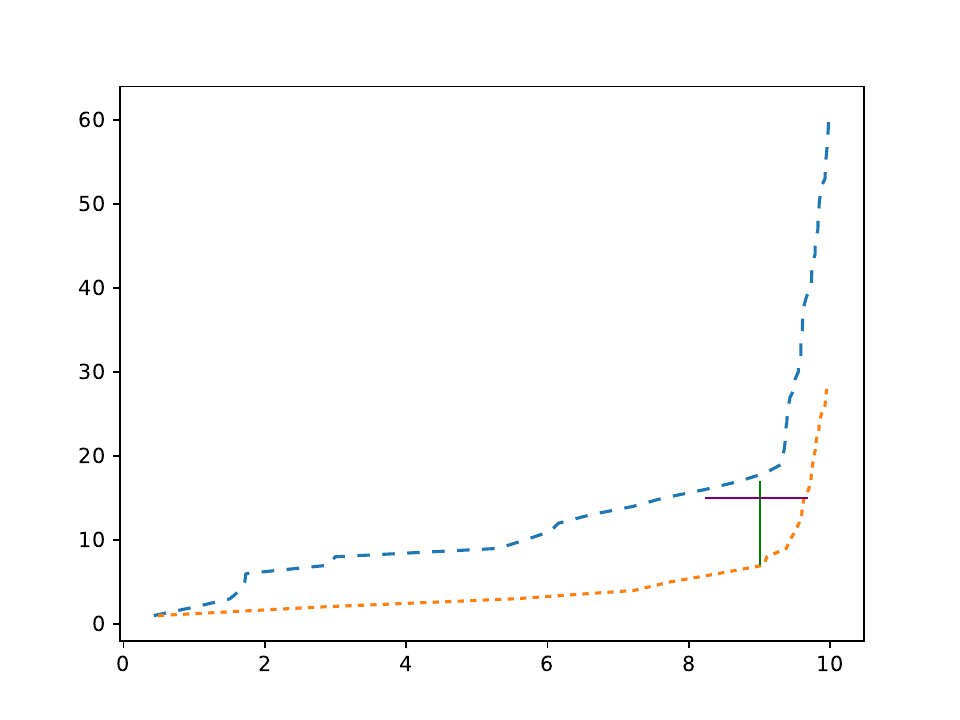}
    \label{fig:criss_cross_cumulative:s3:sclp}}

    \caption{Cumulative input/output diagram for one realization of a criss-cross network.}
    \label{fig:criss_cross_cumulative}
\end{figure*}

\begin{table}[bp]
    \centering
    \caption{Sample results of a single simulation of a criss-cross-network}
    \begin{tabular}{ccc}
        & Autoscaling & Fluid \\
        \hline
        Holding cost & 11556 & 6918 \\
        Response time & 68 & 54 \\
        Failures & 3.26 & 1.04 \\
    \end{tabular}
     \label{tab:criss_cross_example_results}
\end{table}

Table \ref{tab:criss_cross_example_results} shows the results of a single simulation of the criss-cross network over a time period of $10$ units,
in which the fluid policy performs better than autoscaling in all of holding cost, response time, and number of failures.
Figure \ref{fig:criss_cross_cumulative} illustrates the behavior of the 3 individual function types.
The top row contains the values for the autoscaling policy, and the bottom row contains the fluid policy.
Each column represents the cumulative arrivals and departures for one function type.
In each plot,
the x-axis shows time while the y-axis shows counts.
The blue dashed line shows the cumulative count of item arrivals (function calls) and the orange dashed line the cumulative completions.
At any particular time,
the difference between the blue line and the orange line shows the number of concurrent requests in the system,
as illustrated by the green solid line.
At any y-value,
the difference in the x-direction represents the sojourn time of that item (function call) in the system.
This is illustrated by the purple solid line.



\subsection{Results by network size}
\label{sec:comparison_size}
\begin{table}[tbp]
    \centering
    \caption{SCLP vs. Autoscaling by network size}
    \label{tab:results:network-size}
    \begin{tabularx}{\linewidth}{HXH|XXX|XXX}
	    & & & \multicolumn{3}{c|}{Autoscaling} & \multicolumn{3}{c}{Fluid model} \\
    Experiment & Function types & SCLP Obj & Cost & Avg time & Failed & Cost & Avg time & Failed \\
    \hline
    Network size & $50$ & 143760 & 144250 & 2.07 & 21200 & 79697 & 1.14 & 20924 \\
    Network size & $100$ & 284432 & 285195 & 2.06 & 42128 & 163885 & 1.17 & 38570 \\
    Network size & $150$ & 432884 & 429188 & 2.06 & 63340 & 244882 & 1.15 & 59155 \\
    Network size & $200$ & 589762 & 574108 & 2.06 & 84612 & 327981 & 1.15 & 78944 \\
    Network size & $250$ & 739428 & 717340 & 2.06 & 105857 & 423241 & 1.19 & 95929 \\
    Network size & $300$ & 903231 & 865285 & 2.06 & 127762 & 502834 & 1.17 & 118132 \\
    Network size & $350$ & 1052189 & 1009610 & 2.06 & 148935 & 589383 & 1.17 & 136974 \\
    Network size & $400$ & 1195907 & 865284 & 2.06 & 170143 & 675270 & 1.15 & 154336 \\
    Network size & $450$ & 1340775 & 1296753 & 2.06 & 191077 & 759497 & 1.18 & 174226 \\
    Network size & $500$ & 1501960 & 1441829 & 2.06 & 212890 & 838400 & 1.17 & 196228 \\
    \end{tabularx}
\end{table}

In this experiment, we investigate the impact of network size on the relative performance of the fluid model and autoscaling. We vary the number of servers from 10 to 100, with 5 function types per server, resulting in a total of 50 to 500 function types.

The results, as presented in Table \ref{tab:results:network-size}, reveal interesting findings. The average response time remains consistent across different network sizes, indicating that it is not significantly influenced by the number of servers. However, both the holding cost and the number of failures scale linearly with the number of servers.

In each case, the fluid model outperforms autoscaling in all three performance measures. The holding cost and average response time exhibit approximately $75\%$ improvement with the fluid model compared to autoscaling. Moreover, the fluid model exhibited consistently fewer failures, although the improvement was relatively minor.


\subsection{Results by request timeout}

\begin{table}[bp]
    \centering
    \caption{SCLP vs. Autoscaling by timeout}
    \begin{tabularx}{\linewidth}{HXHX|XXX|XXX}
	    & & & & \multicolumn{3}{c|}{Autoscaling} & \multicolumn{3}{c}{Fluid model} \\
    Experiment & Time out & SCLP Obj & Solution time & Cost & Avg time & Failed & Cost & Avg time & Failed \\
    \hline
    Timeout & $2$ & 7507 & 1.28 & 10171 & 0.86 & 0 & 8760 & 0.71 & 85 \\
    Timeout & $5$ & 58251 & 5.97 & 109870 & 2.94 & 392 & 79313 & 1.87 & 190 \\
    Timeout & $10$ & 130940 & 10 & 283169 & 4.91 & 321 & 187093 & 2.75 & 39 \\
    \end{tabularx}
    \label{tab:results:timeout}
\end{table}

In this experiment, we investigated the impact of the timeout value on the relative performance of the fluid model versus autoscaling. Using the methodology outlined in Section \ref{sec:gen_model}, we tackled the SCLP problem with constraints specified in \ref{eqn:gen_timeout}, employing timeout values of 2, 5, and 10 time units.

It is important to note that the resulting SCLP were infeasible for the entire time interval, but instead provided a solution for the maximum time period indicated in Table \ref{tab:results:timeout}. Consequently, the simulations were executed solely for this duration.

The timeout value directly influenced the maximum number of concurrent requests, which we incorporated into the simulator based on constraint \ref{eqn:gen_timeout}, for both the fluid model and autoscaling configurations.

The outcomes of the experiments are presented in Table \ref{tab:results:timeout}. For the shortest timeout value of 2 time units, the fluid model exhibited superior performance in terms of average response time and holding cost, achieving respective improvements of $20\%$ and $15\%$. However, it fared worse in the number of failures metric. One possible explanation for this observation could be the relatively limited feasible region of the fluid approximation, which allows little room for the stochastic variation of the queue.

In contrast, for timeout values of 5 and 10 time units, the fluid model demonstrated significant improvements across all performance measures. It achieved a notable reduction of $60-80\%$ in average response time compared to autoscaling, accompanied by a substantial decrease in the number of failures.

\subsection{Results by initial number replicas}


In this experiment, we aimed to examine the impact of different values of the initial number of replicas in the autoscaler on the trade-off between holding cost and response time, and compare the results to the fluid model.

\begin{table}[htbp]
    \centering
    \caption{SCLP vs. Autoscaling by initial replicas}
    \begin{tabular}{HcH|ccc}
    Experiment & Initial replicas & SCLP Obj & Cost & Avg time & Failed\\
    \hline
    Initialization heuristics & $5$ & 200620 & 144073 & 2.06 & 21215 \\
    Initialization heuristics & $10$ &  & 128846 & 1.83 & 21216 \\
    Initialization heuristics & $15$ &  & 121138 & 1.72 & 21348 \\
    Initialization heuristics & $20$ &  & 119476 & 1.69 & 21109 \\
    Initialization heuristics & $30$ &  & 119638 & 1.69 & 20557 \\
    Initialization heuristics & $40$ &  & 120243 & 1.70 & 20188 \\
    Initialization heuristics & $50$ &  & 120712 & 1.70 & 19805 \\
    Initialization heuristics & Fluid & 200620 & 79626 & 1.14 & 20963 \\
    \end{tabular}
    \label{tab:results:initial-replicas}
\end{table}

We varied the initial number of replicas per function type in the autoscaler between 5 and 50. The results are presented in Table \ref{tab:results:initial-replicas}. The findings indicate that as the initial number of replicas increases, there is an initial improvement in both holding cost and average response time. However, this improvement eventually reaches a plateau and falls short of the performance achieved by the fluid model. Although the number of failures exhibits a slight improvement, transitioning from slightly worse than the fluid model to slightly better, the improvement is marginal.

\begin{figure}[htbp]
\centering
    \subfloat[5 initial replicas, service 1]{\includegraphics[width=0.48\linewidth]{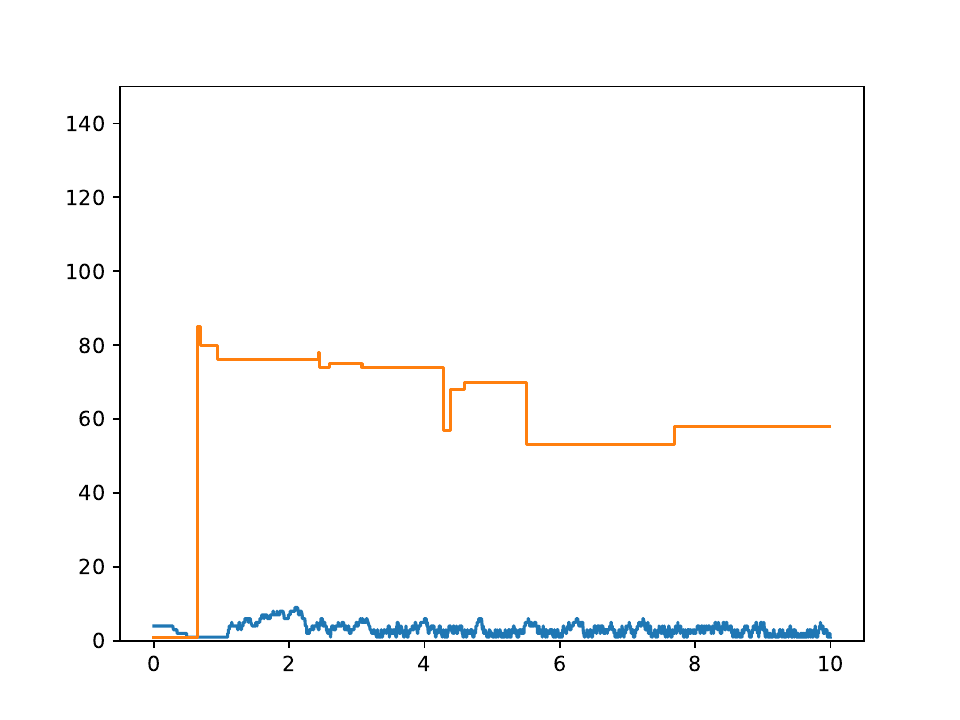}
    \label{fig:num_replicas:10:0}}
    \subfloat[5 initial replicas, service 4]{\includegraphics[width=0.48\linewidth]{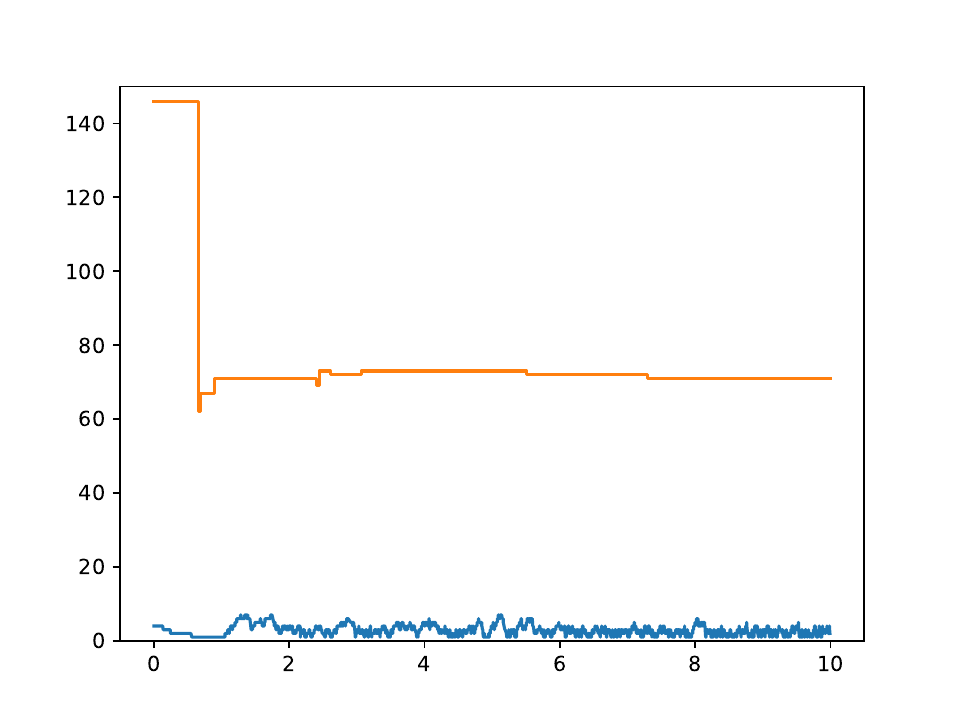}
    \label{fig:num_replicas:10:3}}
    \hfil

    

    \subfloat[50 initial  replicas, service 1]{\includegraphics[width=0.48\linewidth]{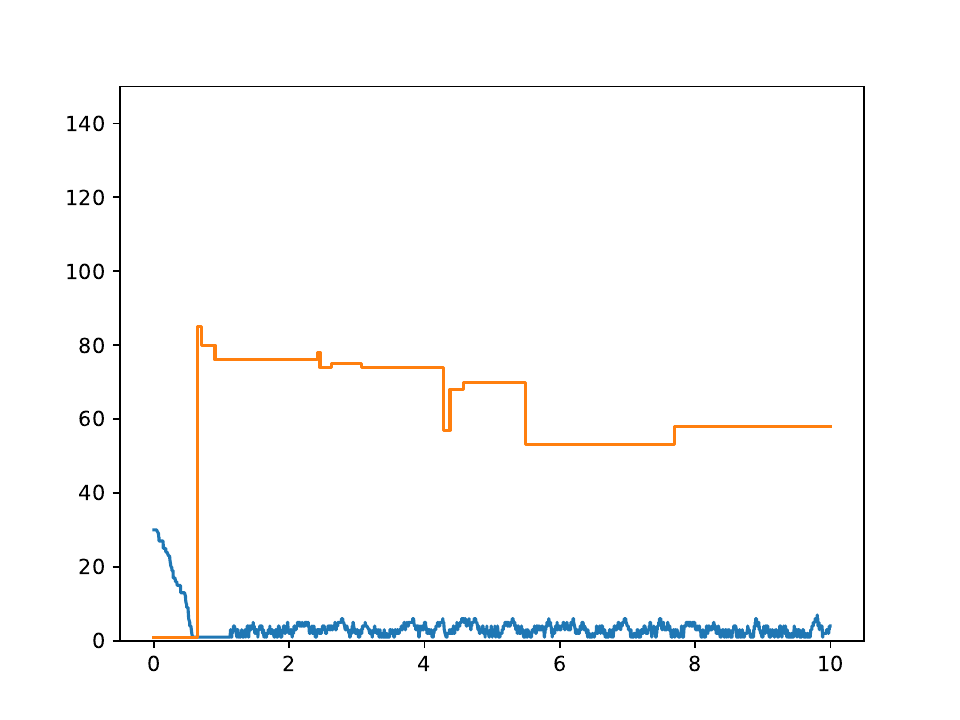}
    \label{fig:num_replicas:100:0}}
    \subfloat[50 initial replicas, service 4]{\includegraphics[width=0.48\linewidth]{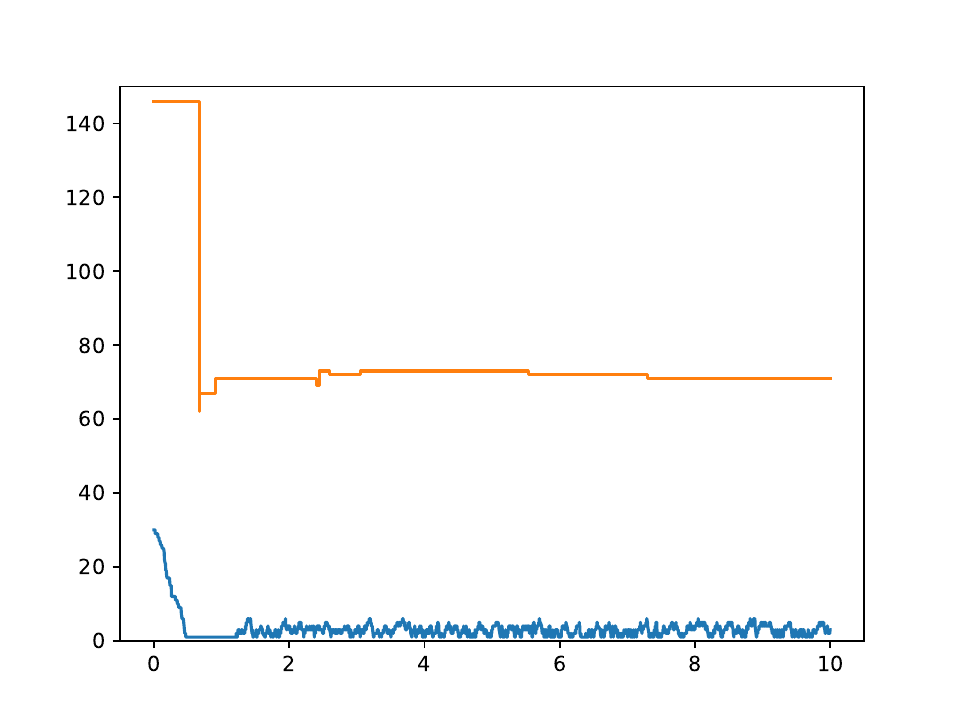}
    \label{fig:num_replicas:100:3}}
    
    \caption{Replicas over time.}
    \label{fig:num_replicas}
\end{figure}

To further illustrate the behavior under the different policies, Figure \ref{fig:num_replicas} depicts the number of replicas over time for two function types during a single simulation, comparing the scenarios of 5 and 50 initial replicas for the autoscaler with the fluid model. The orange lines represent the fluid model, while the blue lines represent autoscaler. It can be observed that the fluid model has the flexibility to increase the number of replicas until the server capacity is fully utilized. In contrast, autoscaling does not maintain a large number of replicas as the failure rate is not high. Consequently, the average response time achieved by the fluid model is significantly better than that of autoscaler.

\subsection{Results for heterogeneous functions}
\label{sec:comparison_hetero}
\begin{table}[bp]
    \centering
    \caption{SCLP vs. Autoscaling by heterogeneity of functions}
    \begin{tabularx}{\linewidth}{HXH|XXX|XXX}
	    & & & \multicolumn{3}{c|}{Autoscaling} & \multicolumn{3}{c}{Fluid model} \\
    Experiment & Rate spread & SCLP Obj & Cost & Avg time & Failed & Cost & Avg time & Failed \\
    \hline
    Heterogeneous functions & $0$ & 127196 & 141422 & 2.08 & 20659 & 80415 & 1.19 & 17692 \\
    Heterogeneous functions & $2$ & 231097 & 163377 & 1.20 & 44251 & 131081 & 0.93 & 29696 \\
    Heterogeneous functions & $5$ & 334574 & 221774 & 0.95 & 76499 & 194387 & 0.79 & 42924 \\
    Heterogeneous functions & $10$ & 408052 & 208885 & 0.62 & 125334 & 241527 & 0.71 & 56154 \\
    \end{tabularx}
    \label{tab:results:heterogeneous}
\end{table}

In this experiment, we investigate the impact of function heterogeneity on the relative performance of the fluid model compared to autoscaling. For each function, we randomly sample the arrival and processing rates independently and uniformly from the range $[100, 100 + 2.1x]$, where $x$ represents  {\em a range spread} value that indicated in the corresponding column of Table \ref{tab:results:heterogeneous}.

The results indicate that the fluid model demonstrates significantly greater robustness against failures compared to autoscaling. The number of failures increases at a slower rate for the fluid model, while autoscaling experiences a much faster growth in failures. Autoscaling failures start at a modest $16\%$ higher for homogeneous function types (with a rate spread of 0) and escalate to a substantial $320\%$ worse for the most heterogeneous case.

On the other hand, the holding cost and response time exhibit improvement with an increasing rate spread. This improvement is more pronounced for autoscaling, ultimately surpassing the performance of fluid model. We propose that this outcome is a consequence of the large number of failures in the autoscaling policy, primarily caused by the uniform configuration employed to handle a heterogeneous load. Manual configuration of autoscaling rules for each function type may alleviate this issue.

%% file: src/relatedwork.tex
\section{Related Work}\label{sec:relatedwork}


In recent studies, significant attention \cite{Baldini2017} has been devoted to the exploration of various serverless platforms that have surfaced from industry, academia, and open-source contributions. OpenLambda, an open-source solution for the development of sophisticated web services and applications within the serverless computing environment, is presented in \cite{Hendrickson2016}. A new performance-focused serverless computing platform, constructed in .NET and deployed on Microsoft Azure using Windows containers for function execution environments, is detailed in \cite{Mcgrath2017}. Sequoia \cite{Tariq2020} is a drop-in front-end for serverless platforms that allows policies to dictate how or where functions should be prioritized, scheduled, and queued. The ideal performance for a specific workload is achieved by carefully designing and evaluating several scheduling algorithms, such as resource-aware scheduling and explicit priority-based scheduling. Atoll \cite{Singhvi2021} is a delay-sensitive serverless framework that exploits a shortest-remaining-slack-first algorithm for scheduling serverless functions. Atoll uses a threshold-based resource scaling method based on queuing delays.


In addition, there is more research work on managing resources or guarantee performances for serverless platforms~\cite{Ali2018,Saha2018,Schuler2020,Shahrad2020}.
The latency-utilization trade-off in serverless platforms is explored in~\cite{Qiu2021}, using IBM Cloud and a private cloud data to study how workload consolidation, queuing delay, and cold starts affect end-to-end function request latency.

Even though resource allocation is managed by the serverless platform, users still have to provide resource requirements for their functions.
This may be done through profiling, yet under- or over-allocation is possible.
\cite{Eismann2021} provides a method for estimating the optimal resource size for serverless functions using monitored data.


ENSURE \cite{Suresh2020} is another rule-based function resource manager.
It allocates $R+c\sqrt{R}$ containers to a function with load $R$, scales the resources within an invoker based on a latency degradation threshold, and scales the number of invokers based on a memory capacity threshold, tuned per function and per workload. The open-source HPA is a simplified version of heuristics-based horizontal autoscaling for containers and it scales out an application based on a customized metric.


To build serverless applications and avoid provider lock-in, Knative was proposed as an open-source, unified Kubernetes serverless API platform~\cite{Kaviani2019}.
A method for determining the number of pods needed to run functions, based on past and future predictions, is provided in~\cite{Fan2020}.


A potential means to model the simultaneous execution of applications, each consisting of running a chain of serverless functions is multichain queueing networks.
In such networks, the service stations correspond to functions, a chain corresponds to the sequence of functions that are executed in order to fulfill an application request, and customers correspond to requests.
The analysis of such networks may be tractable in some cases, using mean-value analysis~\cite{Reiser1980}, or using approximate analysis~\cite{Hsieh1988}.
Autoscaling involves the optimal determination of resource capacity and the number of servers at the service centers, given the load and some global constraints.
This may be achieved by using a combination of an optimizer and a multichain queueing network analyzer.



%% file: src/conclusion.tex
\section{Conclusion}\label{sec:conclusion}
We have addressed the challenge of dynamic autoscaling in serverless cloud computing by proposing a resource allocation approach that adjusts computational resources based on changes in demand. Our model, built upon the MCQN framework, supports function chains, diverse function types, different resource types, and captures the network state within the serverless ecosystem. By employing a fluid approximation of the MCQN model, we formulated the SCLP problem to obtain an optimal dynamic resource allocation policy. The optimal resource allocation is then transformed into the optimal resource allocation per replica and determines the optimal number of replicas within fixed time intervals.

Our approach is capable of proactively finding an asymptotically optimal control policy for large-scale problems in a reasonable timeframe. Moreover, it allows for the recomputation of the optimal fluid policy at desired intervals, enabling adaptation to changing function demand.

To evaluate our approach, we developed a simulation framework that facilitated the analysis of serverless function network models. Through simulation experiments, we compared our method to a simple auto-scaling approach, which adjusts the number of replicas based on load balancer failures or the detection of idle replicas. Performance measures such as holding cost, average response time, and the number of failures and timeouts were assessed. Our method demonstrated superior performance across all measures, although there were instances where it outperformed in average response time and holding cost but underperformed in the number of failures.

Future research directions include conducting additional simulation experiments to cover larger problem sizes. Moreover, there is a need to consider more realistic serverless scenarios and compare our method with existing autoscalers in real-world systems. By addressing these areas, we can further validate the effectiveness and applicability of our approach in practical serverless environments. \ES{would we like to improve this and add more feature work?}